\documentclass[preprint]{aastex}

\received{2000 June 22}
\accepted{2000 September 22}
\slugcomment{accepted for publication in the {\it Astrophysical Journal}}
\shorttitle{Loewenstein, Valinia, \& Mushotzky}
\shortauthors{{\it RXTE} Investigation into
Extended Hard X-ray Emission from Elliptical Galaxies}

\begin{document}

\title{{\it RXTE} Investigation into Extended Hard X-ray Emission from
Elliptical Galaxies} 
\author{Michael Loewenstein\altaffilmark{1}, Azita Valinia\altaffilmark{1},
and Richard F. Mushotzky}
\affil{Laboratory for High Energy Astrophysics, NASA/GSFC, Code 662,
Greenbelt, MD 20771} 
\altaffiltext{1}{Also with the University of Maryland Department of 
Astronomy} 
\email{loew@larmes.gsfc.nasa.gov}

\begin{abstract}
We present results of the first {\it RXTE} investigation of hard X-ray
emission from normal elliptical galaxies. Joint spectral analysis of
{\it RXTE} PCA and {\it ASCA} GIS data for NGC 4472 and NGC 4649
reveals the presence of $\sim 5\ 10^{-12}$ erg cm$^{-2}$ s$^{-1}$ of
2-10 keV emission emanating from a region outside of the optical
galaxy, and of a spectral hardness intermediate between the galactic
hot interstellar gas and X-ray binary components. A large fraction of
this emission originates in the Virgo intracluster medium (ICM) and we
present constraints on the ICM temperature and metallicity in the
vicinity of these two galaxies. However, the diffuse flux is
significantly greater than expected based on the soft X-ray intensity
measured with {\it ROSAT}, allowing for the possible presence of an
additional diffuse non-thermal component.  We also derive new
constraints on the integrated X-ray binary emission from these two
galaxies, and on the presence of a nuclear flat-spectrum point source.
\end{abstract}



\section{Background}

\subsection{Introduction: X-ray Emission From Elliptical Galaxies}

Observations with the {\it Einstein} Observatory IPC found elliptical
galaxies to be strong emitters of X-rays; and, a rich diversity and
complexity of spatial and spectral characteristics were subsequently
revealed by observations with the {\it ROSAT} and {\it ASCA}
satellites.  The emission from the most luminous galaxies is dominated
by hot gas in hydrostatic equilibrium in a galactic potential that
includes dark matter. Density, temperature, and metallicity
distributions in the hot gas derived from these data enabled
investigators to measure the dark matter content in these galaxies and
constrain the dynamical and chemical evolution of their gaseous and
stellar constituents \citep{Mu94,Lo94}.

In addition to emission from 0.5--1 keV gas, an observable hard X-ray
component from the ensemble of galactic X-ray binaries is expected.
The presence of such a component was inferred from the IPC data
\citep{CFT,KFT}, and was first directly detected in BBXRT observations
of NGC 1399 and NGC 4472 \citep{Se93}. {\it ASCA} observations first
revealed the hard component to be extended in some galaxies
\citep{Ma94}.

The nature of the hard component is of fundamental interest for a
number of reasons.  If it originates solely from X-ray binaries, the
hard component should scale as the optical light (both within
individual and among different galaxies) with the same proportionality
observed in the bulges of spiral galaxies.  If the correlation is not
linear, or if the ratio of hard X-ray to optical luminosity differs in
magnitude or dispersion from the spiral bulge value, a fundamental
difference between the stellar populations of bulges and ellipticals
may be indicated. This would have important implications for, e.g.,
the Type Ia supernova rate and the formation of ellipticals.
Moreover, since the depth of the potential well (the velocity
dispersion), and hence the ability to retain hot gas, decreases with
optical luminosity, the hard component should be increasingly
important for fainter galaxies, ultimately dominating below some
luminosity that depends on -- and therefore can help constrain -- how
the dark matter content and Type Ia supernova rate scale with
luminosity \citep{Ci91}.  Additional sources of hard component
emission might originate from a low-luminosity active nucleus
associated with dynamically-inferred supermassive black holes, or from
inverse Compton scattering of the cosmic microwave background and/or
nonthermal bremsstrahlung associated with an extended halo of high
energy electrons. Finally, understanding the hard component can help
resolve some of the ambiguities inherent in current fitting of {\it
ASCA} spectra that are the source of a systematic uncertainty in the
hot gas metallicity.

\subsection{Previous Results on the Hard Component and Motivation for
{\it RXTE} Observations}

Analysis of {\it ASCA} data allowed spectral decompositions into soft
(hot gas) and hard (X-ray binary) components for a large number of
elliptical galaxies for the first time.  \citet{MOM} derived hard
component fluxes for 27 early-type galaxies by fitting {\it ASCA}
spectra, extracted from circular regions enclosing four (optical)
half-light radii $4R_e$ ({\it i.e.}, $\approx 85$\% of the optical
light), with models composed of thermal plasma and 10 keV thermal
bremsstrahlung emission. (The temperature of the latter is poorly
constrained, although a lower limit of 2 keV can be placed; Matsumoto
et al. 1997.) In the 0.5-10.0 keV band, the flux emitted by the hard
component inside these radii (3--8.5$^\prime$) was estimated to range
from $\sim 0.1$--$1.6\ 10^{-12}$ erg cm$^{-2}$ s$^{-1}$ (see, also,
Matsumoto et al. 1997).  Although many galaxies roughly follow the
expected linear trend, there are many cases of excess emission. Based
on examination of the {\it ROSAT} image, a low-luminosity AGN is the
likely explanation for the most extreme case, IC 1459, while extended
emission possibly associated with the Virgo (Fornax) cluster is a
possible contributor in NGC 4472 (NGC 1399).  \citet{BF98} argued for
the presence of an additional intermediate temperature ($\sim 1.5$--2
keV) component to replace or augment the X-ray binary component that
leads to higher abundances in spectral fits.  Using a similar
approach, \citet{ADF} inferred the presence of a hard component
described by a flat power-law originating from a radiatively
inefficient accretion disk around a supermassive black
hole. \citet{Fu99} presented evidence for an extended inverse Compton
or non-thermal bremsstrahlung component in the X-ray emission from
galaxy groups (that otherwise strongly resemble the extended emission
around some elliptical galaxies), similar in nature to that observed
in some clusters of galaxies \citep{SK00}.

Identification and analysis of possible hard X-ray emission from
extended thermal, extended non-thermal, or nuclear non-thermal
components using {\it ASCA} are severely limited by the rapidly
declining effective area above 2 keV, the broad PSF, and dilution from
the ubiquitous presence of hot interstellar gas. Three non-active
ellipticals -- NGC 1399, NGC 4472, and NGC 4636 \citep{Ik92,Aw91} --
were observed with the {\it Ginga} LAC that has a larger $>2$ keV
collecting area.  Initial (pre-{\it ASCA}/{\it ROSAT}) studies fit the
2-10 keV spectra with one-component models, and the higher
temperatures and 2-10 fluxes as compared to subsequent {\it ASCA} and
{\it ROSAT} results are suggestive of the presence of a significant
extended hard component.

We re-analyzed the background-subtracted {\it Ginga} top-layer spectra
for these galaxies, extracted from the LEDAS data base.  Two-component
fits including a soft component determined by the {\it ASCA} spectra
require a strong hard component with 2-10 keV fluxes of $\sim 4$, 12,
and $3\ 10^{-12}$ erg cm$^{-2}$ s$^{-1}$ for NGC 1399, NGC 4472, and
NGC 4636, respectively.  The hard component (thermal bremsstrahlung)
temperatures of 5--20 keV for NGC 4472 and NGC 4636 and 2--5 keV for
NGC 1399 are compatible with the {\it ASCA} spectra; however, the
fluxes are well in excess (by a factor of $\sim 3$) of what is
formally allowed in the {\it ASCA} fits. (For NGC 1399, the BBXRT
estimate of the hard component flux is closer to the {\it Ginga} than
the {\it ASCA} estimate.)

How might the {\it ASCA} and {\it Ginga} fluxes be reconciled?
Perhaps systematic background uncertainties that can be considerable
for {\it Ginga} and at high energies for {\it ASCA} are responsible --
note that the {\it Ginga} background subtraction quality is deemed
acceptable only in the case of NGC 1399. Or perhaps the hard component
is more extended, or has a harder and/or more complex spectrum than
anticipated.

{\it RXTE} observations can greatly facilitate our understanding of
this poorly understood hard component in elliptical galaxies because
of the large effective area of the Proportional Counter Array (PCA) in
the 2-20 keV band ($\sim 1300$ cm$^2$ for each of five Proportional
Counter Units, PCUs). With the recent improvement in background
modeling, PCA spectra (alone, and combined with {\it ASCA} spectra)
can be analyzed to confirm or disconfirm the tentative {\it Ginga}
results and, if the former, distinguish among different emission
mechanisms and constrain spectral model parameters. For our initial
study, representing the first {\it RXTE} observations of elliptical
galaxies, we observed two nearby systems for $\sim 50000$ s.

These Virgo Cluster galaxies are NGC 4472, that requires the most
luminous hard component in fits to both {\it ASCA} and {\it Ginga}
spectra; and, NGC 4649, with the second largest {\it ASCA} hard
component flux and a compact X-ray morphology that distinguishes it
from the {\it Ginga}-observed systems. Both galaxies were observed and
extensively studied with {\it ROSAT} and {\it ASCA}. The upper limit
to any nuclear point source in NGC 4472 (NGC 4649) from {\it ROSAT}
HRI analysis is $\sim 0.0025$ ($\sim 0.005$) HRI cts s$^{-1}$.  No
point sources with 2-10 keV fluxes exceeding $3\ 10^{-13}$ erg
cm$^{-2}$ s$^{-1}$ (converted from HRI count rates assuming a slope
1.7 power-law) are present in the PSPC fields of view of either
galaxy.

\section{Data Analysis}

\subsection{Description of Observations}

NGC 4472 was observed at the end of {\it RXTE} Epoch 3 (16 -- 22 March
1999) for 49888 s, NGC 4649 at the beginning of Epoch 4 (27 March -- 4
April 1999) for 49616 s. We consider only spectra from the PCA
detectors. For most of the NGC 4472 observation, four of the five PCUs
(PCU0, PCU2, PCU3, PCU4) were turned on; the NGC 4649 observation was
primarily performed with either 3 (PCU0, PCU2, PCU3) or two (PCU0,
PCU2) on.

\subsection{Data Reduction and Spectral Extraction}
 
Data reduction is performed with {\it RXTE}-specific FTOOLS and
scripts, following standard procedures recommended by the {\it RXTE}
Guest Observer Facility for faint sources\footnote
{http://heasarc.gsfc.nasa.gov/docs/xte/xhp\_proc\_analysis.html}. A
list of (non-slew) PCA Standard 2 science array files is compiled, and
filter files are created for each corresponding observation
identification and subsequently merged. Good time intervals are
defined using these files in order to exclude data taken with
elevation angle with respect to the bright Earth less than 10 degrees,
with offset from the nominal pointing direction greater than 0.02
degrees, within 30 minutes of the the peak of the last SAA passage, or
with high electron contamination to the background.  Pulse-height
spectra are then extracted by applying this time-filtering for the sum
of all PCU Xenon layers and for the top layer only, restricted to the
most frequent PCU configurations. The final exposure times for the
extracted spectra are 35040 s for NGC 4472, and 31952 (10656) s for
the 3 (2) PCU configuration for NGC 4649. Spectral response matrices
are generated with the FTOOLS v5.0 PCA Response Matrix Generator.
For NGC 4649, a total (42608 s) spectrum is created by summing 3- and
2- PCU spectra weighted by exposure, and the corresponding response
matrix by a count-rate-weighted sum.

Background files for each observation interval are created from the
most recent (February 2000) faint source models corresponding to the
appropriate epoch, and background spectra are extracted with identical
selection criteria and binning to the total (source$+$background)
spectra extracted from the data as described above. Figures 1a and 1b
illustrate how well the background models reproduce the high energy
(top layer) data. For NGC 4472, 88500 $E>2$ keV background-subtracted
counts are detected (about four times the confusion limit), 90\% in
the 2-10 keV band. For NGC 4649, 59650 $E>2$ keV source counts are
detected, with no measurable contribution originating above 15 keV.

\subsection{Data Analysis Scheme}

We fit the {\it RXTE} PCA spectra both individually and simultaneously
with {\it ASCA} GIS spectra -- of the {\it ASCA}/{\it ROSAT}
detectors, the GIS has the most significant bandpass overlap with {\it
RXTE} and is used to constrain the $<10$ keV X-ray properties on the
optical galaxy scale. The effective apertures of the PCA and GIS are
120 and 50 arcminutes in diameter, respectively; although, only a
fraction of the GIS field-of-view (fov) is utilized.

Following \citet{Wh00}, 1-10 keV spectra are extracted for both GIS
detectors within apertures of radius $6R_e$, a radius that encloses
$\sim 90$\% of the optical -- and presumably $\sim 90$\% of the
integrated X-ray binary -- luminosity\footnote{we use the second of
two 1993 NGC 4472 observations ({\it ASCA} sequence id 60029000);
there is an additional observation with the galaxy further
off-axis}. GIS exposure times after screening, and (local-background
subtracted) count rates per detector in these apertures are 23 ks and
0.18 s$^{-1}$ for NGC 4472, and 41 ks and 0.06 s$^{-1}$ for NGC
4649. Local background spectra are extracted from annuli centered on
the galactic nuclei, extending from $12-16'$ ($10-15'$) in the GIS2
(GIS3) detector for NGC 4472, and from $7-11'$ ($7-10'$) in the GIS2
(GIS3) detector for NGC 4649 (blank-sky background spectra have also
been utilized; see Section 5.1).  These spectra are simultaneously fit
with two-component models consisting of (Raymond-Smith) thermal plasma
(the ``ISM'' component) and thermal bremsstrahlung (the ``binary''
component) emission, absorbed by the Galactic column density (see
Section 3.1).  Whether (and, in what sense) such a simple model is
correct is not our concern here; we merely require an accurate
physical characterization of the galactic emission in order to
investigate any additional, presumably extended, hard component
detected by {\it RXTE}.

The {\it RXTE} PCA spectra are independently fit in order to find the
optimum selection criteria (bandpass, total or top-layer spectrum) and
estimate the total hard X-ray flux. Our final physical constraints are
the product of joint two- and three-component PCA/GIS spectral
fitting, conducted as follows.

The two-component models are simple generalizations of the
ISM$+$binary model used to characterize the GIS spectra, with the
spectral model parameters (temperatures, etc.) assumed identical for
the GIS and PCA spectra but the normalizations allowed to
independently vary. That is, any emission extending beyond the GIS,
and into the PCA, fov is assumed to be a simple extension of the
galactic ($<6R_e$) emission.

In the three-component models the ISM and binary normalizations are
assumed equal for the PCA and GIS spectra, as well. (Slightly higher
normalizations, as expected given its larger fov, in the PCA can be
accommodated but do not significantly improve the quality-of-fit when
an additional component is included.) However, an extra component (the
``extended'' component) is now included that is assumed to
significantly contribute only to the PCA spectrum, {\it i.e.} it is
assumed to be a distinct hard component extended on the super-galactic
scale. This is modeled as either a power-law or an additional thermal
plasma. We now discuss the results of these fits; details are
displayed in Tables 1 and 2.

All quoted uncertainties correspond to $\Delta\chi^2=2.7$.

\section{Data Analysis Results: NGC 4472}

\subsection{Review of Recent X-ray Analysis}
There are numerous published results of X-ray analyses of both {\it
ROSAT} \citep{Fo93,DW96,IS96,BB98,IS98,Be99} and {\it ASCA} (Matsumoto
et al. 1997; Buote \& Fabian 1998; Allen et al. 2000; Matsushita et
al. 2000) observations of NGC 4472, as well as the {\it Ginga}
analysis of \citet{Aw91}. Integrated PSPC spectra are characterized by
$kT\approx 1$ keV, roughly solar abundances, and a Galactic column
density ($N_H=1.66\ 10^{20}$ cm$^{-2}$).  The {\it ASCA} spectra
require extra absorption ($0.3-1.5\ 10^{21}$ cm$^{-2}$) and an
additional hard component that can be modeled as $kT\sim 10$ keV
thermal bremsstrahlung emission or a power-law with photon index $\sim
1.8$. Fits to integrated {\it ASCA} spectra yield $kT\approx 0.9$ keV
and roughly one-third solar abundances, both significantly lower than
in PSPC fits.  If the hot gas is modeled by a two-temperature thermal
plasma, an additional power-law component with a flatter photon index
and more nearly solar abundances can be accommodated.  Using the
best-fit models to convert to a common 0.5-2.0 keV bandpass, and the
surface brightness profile from \citet{FJ00} to normalize to a common
aperture of $10.4^{\prime}$ ($6R_e$), we find that published estimates
correspond to a flux of $\sim 9\ 10^{-12}$ erg cm$^{-2}$ s$^{-1}$,
with some of the {\it ASCA} values coming in slightly lower -- in part
due to the higher required {\it ASCA} columns (quoted fluxes in this
paper are not corrected for absorption).  Based on examination of a
PSPC image mosaic of the Virgo Cluster provided by S. Snowden, we
estimate that an additional $\sim 3\ 10^{-12}$ erg cm$^{-2}$ s$^{-1}$
of galaxy flux is emitted in the 10-30$^{\prime}$ annulus, but no
significant (galactic) source flux in the 30-60$^{\prime}$ annulus.

The PSPC and {\it ASCA} spectral analysis column density discrepancies
drive us to restrict our GIS analysis to the 1-10 keV bandpass and fix
the column density at the Galactic value (this effects the best-fit
ISM temperature by $<0.1$ keV and has no impact on any harder
component parameters).  Spectral parameter constraints from fits to
GIS spectra extracted from $10.4^{\prime}$ ($6R_e$) radius circular
regions are as follows: ISM temperature $kT_{\rm ISM}=0.90$
(0.80--0.98) keV, ISM metallicity $Z_{\rm ISM}=0.31$ (0.22--0.58)
solar, 0.5-2.0 keV ISM flux $f_{\rm ISM}=9.3\pm 0.6\ 10^{-12}$ erg
cm$^{-2}$ s$^{-1}$, binary temperature $kT_{\rm bin}=6.0$ (3.3--13)
keV, 2-10 keV binary flux $f_{\rm bin}=2.8\pm 0.35\ 10^{-12}$ erg
cm$^{-2}$ s$^{-1}$ -- all generally consistent with previous analyses.

\subsection{The {\it RXTE} Spectrum of NGC 4472}
We fit {\it RXTE} PCA spectra independently in order to find the
optimum selection criteria and estimate the total hard X-ray flux from
NGC 4472. The 2.5-12.5 keV top-layer spectrum maximizes statistical
accuracy, and is well-fit by a simple power-law with photon index
$\Gamma=3.21\pm 0.08$. Other simple single-component models are
unacceptable (reduced $\chi^2$ of 2.6 and 1.8 -- $\Delta\chi^2$ of 50
and 28 relative to the power-law -- for thermal bremsstrahlung and
Raymond-Smith plasma models, respectively). We derive a 2-10 keV flux
of $8.8\pm 0.2\ 10^{-12}$ erg cm$^{-2}$ s$^{-1}$ ($\sim 5$\% of which
originates from two point sources lying to the southwest of the
nucleus), compared to a total ($R<10.4^{\prime}$) 2-10 keV GIS flux of
$3.6\pm 0.18\ 10^{-12}$ erg cm$^{-2}$ s$^{-1}$.  The former lies
between the {\it Ginga} estimate of \citet{Aw91} and our own. In the
following joint PCA/GIS fits (Tables 1 and 2), we evaluate the
possible origins of the additional hard X-ray flux measured with {\it
RXTE}: an extension of the NGC 4472 Galactic emission, diffuse Virgo
intracluster medium (ICM) emission, or some previously undiscovered
source.

For the two-component fits without a distinct extended component, {\it
both} the ISM and binary fluxes must be allowed to exceed their
GIS-measured values in the PCA spectrum in order to obtain an optimal
fit ($\Delta\chi^2\approx 60$ compared to the best-fit where only the
binary normalizations are allowed to differ). While this provides a
formally acceptable fit, the required magnitude of additional ISM
0.5-2.0 keV flux in the {\it RXTE} fov ($>2\ 10^{-11}$ erg cm$^{-2}$
s$^{-1}$) over-produces the PSPC flux by about an order of
magnitude. We conclude that there truly is an extra, extended
component in NGC 4472 measured by {\it RXTE} that has a distinct
spectral shape from the ISM and binary components.

Three-component models, where the third (extended) component
contributes only to the PCA spectrum, provide acceptable fits for
either power-law (reduced-$\chi^2=281/320$) or thermal plasma
(reduced-$\chi^2=279/319$) characterizations of the extended emission.
For the former, the constraints on the extended component are photon
index $\Gamma=3.6$ (3.3--3.8), and 2-10 keV power-law component flux
$f_{\rm pow}=5.3\pm 0.3\ 10^{-12}$ erg cm$^{-2}$ s$^{-1}$; for the
latter (presumably diffuse Virgo ICM emission), a temperature $kT_{\rm
ICM}=1.7\pm 0.2$ keV and metallicity $Z_{\rm ICM}=0.44$ (0.17--0.86)
solar, 2-10 keV ICM flux $f_{\rm ICM}=5.0\pm 0.3\ 10^{-12}$ erg
cm$^{-2}$ s$^{-1}$ . The joint constraints on the binary component
temperature are tighter than from the GIS alone: $kT_{\rm bin}=5.0$
(3.8--6.6) keV for the model with power-law extended component, 6.5
(5.2--9.1) keV for the model with thermal extended
component. Additional absorption is not excluded at the $\sim 2\
10^{22}$ cm$^{-2}$ level for the power-law model -- such a steep
spectrum must turnover at some energy $>2$ keV to avoid conflict with
the GIS and PSPC measurements.

The PCA and GIS spectra and the best three-component model fit,
including a thermal extended component, to the joint dataset are shown
in Figure 2; the unfolded model and its components in Figure 3.  Note
the Fe K feature in the unfolded PCA spectrum that determines the
abundance constraint on the extended component.

\section{Data Analysis Results: NGC 4649}

\subsection{Review of Recent X-ray Analysis}
There are extensive published results of analyses of X-ray
observations of NGC 4649, based on both {\it ROSAT}
\citep{DW96,TFK,BB98,IS98,Be99} and {\it ASCA} (Matsumoto et al. 1997;
Buote \& Fabian 1998; Allen et al. 2000; Matsushita et al. 2000) data.
The consensus from {\it ROSAT} data analysis is that the PSPC ($<2$
keV) spectrum can be characterized by a thermal plasma with $kT\approx
0.85$ keV and half-solar abundances, absorbed by cold gas at the
Galactic column density ($N_H=2.2\ 10^{20}$ cm$^{-2}$). The {\it ASCA}
spectra require extra absorption ($1-2\ 10^{21}$ cm$^{-2}$) and an
additional hard component, but are otherwise consistent with the
PSPC-derived parameters. As with NGC 4472, an additional power-law
component with a flatter photon index and more nearly solar abundances
can be accommodated in models with a multi-temperature thermal plasma.
X-ray fluxes derived from these analyses are discrepant. Converting to
a common 0.5-2.0 keV bandpass and $6R_e$ ($7.3^{\prime}$) aperture as
for NGC 4472, we find that published estimates correspond to a range
$\sim 3-5\ 10^{-12}$ erg cm$^{-2}$ s$^{-1}$.  {\it ASCA} estimates are
generally lower than those from the PSPC, a tendency related to the
higher required {\it ASCA} columns. There is no evidence from these
data for significant X-ray emission extending beyond $6R_e$.

1-10 keV spectra are again extracted for both {\it ASCA} GIS detectors
within $6R_e$ apertures, and local-background subtracted. The GIS
spectra are well-fit by our standard two-component model with $kT_{\rm
ISM}=0.76$ (0.63--0.91) keV, $Z_{\rm ISM}=0.29$ (0.19--0.66) solar,
$f_{\rm ISM}=3.5\pm 0.3\ 10^{-12}$ erg cm$^{-2}$ s$^{-1}$ (0.5-2.0
keV), $kT_{\rm bin}=8.9$ ($>5.6$) keV, $f_{\rm bin}=1.4\pm 0.15\
10^{-12}$ erg cm$^{-2}$ s$^{-1}$ (2-10 keV) -- again consistent with
previous analysis results.

\subsection{The {\it RXTE} Spectrum of NGC 4649}
We find that the co-added {\it RXTE} PCA top-layer spectrum in the
3-15 keV energy band optimizes statistical accuracy. The best-fit
(reduced-$\chi^2=32/25$) simple model for this spectrum consists of an
absorbed power-law with column density $N_H=4.0\pm 1.3\ 10^{22}$
cm$^{-2}$ and photon index $\Gamma=3.17\pm 0.18$. A thermal
bremsstrahlung model provides a significantly worse fit
($\Delta\chi^2=10$), although excess absorption is not required in
this case. Although the background model may systematically
overestimate the true background level at the lowest energies (Figure
1b), fits to spectra with higher energy lower bounds yield the same
power-law slope (while, of course, allowing smaller absorbing columns
and larger 2-10 keV fluxes).  The 2-10 keV flux of $5.9\pm 0.1\
10^{-12}$ erg cm$^{-2}$ s$^{-1}$ is about a factor of four higher than
that measured with the GIS.

In contrast to NGC 4472, a two-component model lacking a distinct
extended component does not provide as good a fit to the joint PCA/GIS
dataset as the three-component model ($\Delta\chi^2\approx 15$, one
additional degree of freedom) and such fits allow equal GIS and PCA
ISM component normalizations (as might be expected from the absence of
significant measurable galactic PSPC emission outside of our GIS
aperture) -- see Table 2.  The binary component spectral parameters
for such a fit are $kT_{\rm bin}=4.2$ (3.9--4.6) keV, $f_{\rm
bin}=1.4\pm 0.1\ 10^{-12}$ erg cm$^{-2}$ s$^{-1}$ (GIS fov, 2-10 keV),
$f_{\rm bin}=6.2\pm 0.15\ 10^{-12}$ erg cm$^{-2}$ s$^{-1}$ (PCA fov,
2-10 keV). The relative binary fluxes indicate a roughly $r^{-2}$
emissivity profile.

Three-component models with an additional component contributing only
to the PCA spectrum favor power-law over thermal plasma
characterizations (reduced-$\chi^2=255/246$ compared to 264/246) --
but only if intrinsic absorption is included in the former.  The
constraints on the extended component are photon index $\Gamma=3.9\pm
0.5$, $N_H=7.1$ (4.6--9.4) $10^{22}$ cm$^{-2}$, $f_{\rm pow}=4.1\pm
0.15\ 10^{-12}$ erg cm$^{-2}$ s$^{-1}$ (2-10 keV) for the power-law;
or, $kT_{\rm ICM}=3.4\pm 0.4$ keV, $Z_{\rm ICM}=0$ ($<0.037$ solar),
$f_{\rm ICM}=4.8\pm 0.2\ 10^{-12}$ erg cm$^{-2}$ s$^{-1}$ (2-10 keV)
for the ICM thermal plasma.  The excess absorption reflects the
requirement that the power-law turn over, but need not be construed as
implying the presence of a true extended column density of cold gas
since this is not apparent from {\it ROSAT}/{\it ASCA} spectral
analysis of the Virgo ICM.  The improved, joint constraints on the
binary component temperature are $kT_{\rm bin}=15$ (8.6--26) keV for
the model with a power-law extended component, 9.6 (6.7--19) keV for
the model with thermal extended component. See Tables 1 and 2, and
Figures 4 and 5 for further details. (Note that the dynamic ranges for
these figures are identical to their NGC 4472 counterparts, Figures 2
and 3.)  In contrast to the case of NGC 4472, an Fe K line is not
required in the fits with an extended thermal component.

\section{Discussion}

\subsection{Summary of Extended Component Constraints and the Virgo
Cluster Contribution}

We have measured additional 2-10 keV X-ray fluxes in the {\it RXTE}
PCA spectra of NGC 4472 and NGC 4649 of $\sim 5\ 10^{-12}$ erg
cm$^{-2}$ s$^{-1}$ above the galactic ($r<6R_e$) emission detected by
the {\it ASCA} GIS (3.8 and $1.6\ 10^{-12}$ erg cm$^{-2}$ s$^{-1}$ for
NGC 4472 and NGC 4649, respectively). This extra emission arises from
an extended component -- the effective PCA aperture subtends a $\sim
10$ (20) times larger solid angle than those used for extracting the
NGC 4472 (NGC 4649) GIS spectra -- and must be spectrally steep at
$E\gtrsim 2$ keV to avoid exceeding the relatively compact, flat hard
component present in GIS, as well as PCA, spectra (see Figures 3 and
5). Note that the extended component contribution to the GIS spectra
is negligible when aperture-scaled and, in any case, is subtracted off
since local background subtraction is employed.

X-ray emission associated with the Virgo intracluster medium must
contribute to the extended hard component of both
galaxies. \citet{IS96} analyzed the cluster emission near NGC 4472,
deriving a 0.2-2.5 keV surface brightness of $2.4\ 10^{-15}$ erg
s$^{-1}$ cm$^{-2}$ arcmin$^{-2}$. (There is emission from the North
Polar Spur in this direction as well; however, this does not
significantly contribute to any of the spectra analyzed here that
exclude energies $<1$ keV.)  The metallicity and temperature were not
well-determined, although the latter is constrained to exceed 1.1
keV. Examination of a PSPC image mosaic of the Virgo Cluster provided
by S. Snowden, yields an intensity about 15\% lower near NGC 4649
(estimated from averaging over various annuli with inner radius
$5-10^{\prime}$ and outer radius $20-60^{\prime}$ measured from the
center of NGC 4649). The spectral parameters in our joint fits for NGC
4472 including an ICM-like extended thermal component are consistent
with those of \citet{IS96}, and we use these to convert their surface
brightness to the 2-10 keV band. This converted PSPC Virgo ICM
emission can account for $\sim 40-75$\% of the extended hard emission
in NGC 4472, and $\sim 40-60$\% in NGC 4649 (Table 3).  We confirm
this by re-analyzing the {\it ASCA} GIS data, substituting blank-sky
(extracted with the FTOOL ``mkgisbgd'') for local background. We
repeat the GIS spectral fits described above (Sections 3.1, 4.1) with
the inclusion of an additional intracluster component with parameters
fixed at the best-fit values obtained from joint PCA/GIS spectral
analysis, and obtain excellent consistency with the PCA fluxes (Table
3).

Several arguments support a thermal intracluster origin for the
extended hard emission. In particular, the extended component spectra
are consistent with what one expects for the Virgo ICM -- especially
for NGC 4472 where an Fe K feature of reasonable strength is
consistent with the data.  Temperatures are in line with previous
measurements \citep{NB95,Ki00}, with the lower temperature for NGC
4472 perhaps reflecting the shallower gravitational potential of the
NGC 4472 sub-cluster rather than the Virgo Cluster at large.  The
similarity of the NGC 4472 and NGC 4649 extended component fluxes also
argues for an ``external'' origin. If, instead, it were associated
with some population of relativistic particles co-existing with the
{\it galactic} thermal plasma, one would expect stronger emission from
NGC 4472 with its more luminous and extended soft X-ray halo: the
total NGC 4472 gas mass is $\sim 100$ times that of NGC 4649
\citep{IS96,KMa97}.

On the other hand, the ICM temperature for NGC 4649 is somewhat higher
than might be expected, and the very low abundance upper limit
inferred from the absence of Fe K emission may be puzzling -- although
abundance gradients have been measured (e.g., Kikuchi et al. 1999)
accurate measurements localized to large radii are rare (the distance
of NGC 4649 from the center of the Virgo Cluster is approximately the
cluster virial radius). Moreover, assuming an accurate {\it
ROSAT}/{\it RXTE} cross-calibration, the 2-10 keV flux is 30--150\%
higher than calculated based on an extension of the PSPC flux and NGC
4472 ICM model (although consistent if the NGC 4649 ICM model is
used). Perhaps there is an additional nonthermal {\it cluster}
component present at modest levels in the PCA spectra of both
galaxies.

\subsection{Comparison of Integrated X-ray Binary Properties}

The combined PCA/GIS dataset enable us to place improved limits on the
parameters of the compact (galactic) hard component -- assumed here to
originate as the integrated emission from galactic X-ray binaries (see
Tables 1 and 2).  If the extended hard component is assumed to be a
power-law for both galaxies, the binary component temperature
constraints are disjoint (Table 1) -- 3.8--6.6 ($>8.6$) keV for NGC
4472 (NGC 4649), but the fluxes (Table 2) consistent when normalized
to the optical luminosities (NGC 4472 is $\sim 40$\%
brighter). Conversely, the binary temperatures are consistent assuming
thermal (ICM) models for the extended components -- 5.2--9.1 ($>6.7$)
keV for NGC 4472 (NGC 4649), but the NGC 4472 X-ray flux exceeds that
of NGC 4649 by a factor of $\sim 2$ (1.4 relative to their respective
optical luminosities).  Thus there are apparent differences in the
binary populations, even though the stellar populations of these
galaxies are nearly indistinguishable \citep{Tr00}. This confirms the
findings of \citet{Wh00} based on {\it ASCA} data alone, who finds a
significant range in the ratio of X-ray binary to optical luminosity
amongst ellipticals in general, and a similar luminosity variation
between NGC 4472 and NGC 4649 to that reported above.  The
X-ray-to-blue luminosity ratio is a factor of 2.0--3.5 (1.5--2.5)
times greater for NGC 4472 (NGC 4649) than the typical values inferred
from earlier {\it ASCA} spectral analysis \citep{HMa97,KMa97}.

An extension of the galactic ($<6R_e$) binary component (as might be
expected from, e.g., an extended halo of globular clusters) can
account for the extended hard component in NGC 4649 without violation
of the measured diffuse PSPC intensity, although models with a
spectrally distinct extended component provide formally superior fits
(see Section 4.2 and Table 1).  However, since this is not the case
for NGC 4472 (with its more prominent globular cluster population and
larger X-ray extent), we consider this an unlikely explanation for the
excess flux detected with the PCA.

\subsection{Constraints on the Presence of a Flat Power-law}

Allen et al. (2000) fit {\it ASCA} spectra with a model consisting of
a two-phase ISM plus a relatively flat (compared to luminous AGN)
power-law, identifying the latter with a new type of low-luminosity,
spectrally hard nuclear X-ray source. Figures 6 and 7 (``$\nu
F_{\nu}$'' plots, as in Figures 3 and 5) show their derived power-law
components (best-fits and 90\% confidence limits) superposed on the
PCA data. At energies below 10 keV, this component would be
overwhelmed by the extended hard emission, but should become
conspicuous at higher energies. The presence of such a component is
ruled out in NGC 4649 by the virtual absence of $>15$ keV PCA counts,
but appears to be present in the binned hard spectrum of NGC
4472. Note, however, that its addition does not improve global
spectral fits and the formal upper limit for such a component is
comparable to what is derived by Allen et al. (2000). Moreover, recent
{\it Chandra} results (Loewenstein et al., in preparation) rule out a
nuclear point source of this magnitude in NGC 4472 unless the emission
below $\sim 10$ keV is cut off by absorption. For NGC 4649, we derive
an upper limit of $1.3\ 10^{-13}$ erg cm$^{-2}$ s$^{-1}$ to the 2-10
keV flux of any additional power-law component with slope equal to the
best-fit value from Allen et al. (2000), $2.4\ 10^{-13}$ erg cm$^{-2}$
s$^{-1}$ if their steepest allowed slope is assumed, compared to their
inferred flux of $6.2\ 10^{-13}$ erg cm$^{-2}$ s$^{-1}$. These are
derived from fits that include PCA energies up to 30 keV; if only the
10-30 keV PCA band is utilized, the former limit is reduced to $8\
10^{-14}$ erg cm$^{-2}$ s$^{-1}$.

\section{Conclusions}

The X-ray emission from the optical portions of elliptical galaxies
includes contributions from both hot interstellar gas and X-ray
binaries.  Using newly obtained {\it RXTE} data we have detected and
investigated an additional component that is both spectrally and
spatially distinct, being more extended than the galactic emission and
of a hardness intermediate between the two galactic components.  The
extended hard component can be modeled as a steep ($\Gamma >3.3$)
power-law in the 2-10 keV band, but is perhaps more naturally
explained as a thermal plasma.

The fact that the ratio of extended hard component flux in NGC 4472 to
that in NGC 4649 is consistent with the ratio of local diffuse Virgo
Cluster soft X-ray intensity suggests that this emission is
predominantly associated with the cluster and not the individual
galaxies (the NGC 4472 galactic emission is $\sim 4$ times brighter),
and the spectral parameters ($kT=1.7\pm 0.2$ keV and $Z=0.17-0.86$
solar for NGC 4472, $kT=3.4\pm 0.4$ keV, $Z=<0.037$ solar for NGC
4649) are reasonable for Virgo intracluster plasma -- though somewhat
extreme for NGC 4649. However, the measured PCA and GIS 2-10 keV
intensities of this diffuse emission are significantly greater than
expected based on the soft diffuse surface brightness measured by the
PSPC, leaving open the possibility of the presence of an additional
nonthermal source of hard X-rays at the $\sim 10^{-15}$ erg cm$^{-2}$
s$^{-1}$ arcmin$^{-2}$ (2-10 keV) level.

We also placed new limits on the temperature and 2-10 keV flux of the
integrated X-ray binary component, although their precise values
depend on the model employed for the extended hard component.  Despite
the similarities of their stellar populations, the binary populations
in NGC 4472 and NGC 4649 are distinct.  If the extended hard component
is modeled as thermal plasma emission, the ratio of X-ray binary to
optical luminosity is a factor of 1.4 larger for NGC 4472 than for NGC
4649.
 
Finally, there is a suggestion of a flat-power law contribution to the
very high energy ($>20$ keV) X-ray spectrum of NGC 4472 at a level
consistent with that suggested by Allen et al. (2000) to arise from
the galactic nucleus; however, such a component is not evident in the
NGC 4649 spectrum.

In this paper, we emphasize the spatial and spectral complexity of the
hard X-ray emission from elliptical galaxies and make some tentative
first steps toward understanding its origin and nature. Further
progress awaits hard X-ray observation of more ellipticals, and
improved hard X-ray imaging data.
 
\acknowledgments

We gratefully acknowledge the support of Keith Jahoda and the {\it
RXTE} GOF to the PCA data reduction effort, and also thank Ray White
and Steve Snowden for assistance with {\it ASCA} GIS and {\it ROSAT}
PSPC data analysis, respectively.  We made extensive use of the High
Energy Astrophysics Science Archive Research Center database, and the
University of Leicester Database and Archive Service at the Department
of Physics and Astronomy, Leicester University, UK. We also thank the
referee, Jimmy Irwin, for his constructive and beneficial comments.

\clearpage


\clearpage

\begin{deluxetable}{cccccccccc}
\tablecaption{PCA/GIS Joint Spectral Fits -- Best-fit Parameters}
\tablewidth{0pt}
\tablehead{
\colhead{Model} & \colhead{$kT_{\rm ISM}$} & \colhead{$Z_{\rm ISM}$} &
\colhead{$kT_{\rm bin}$} & \colhead{$\Gamma$ or $kT_{\rm ICM}$} & 
\colhead{$N_{\rm H,\Gamma}$ or $Z_{\rm ICM}$} & \colhead{${\chi_\nu}^2$} 
}
\startdata
\sidehead{\bf NGC 4472:}
gx & 0.89(0.78--0.97) & 0.42(0.26--0.77) & 4.2(3.6--4.8)                       
   & ...              & ...              & 280/320\\
gp & 0.89(0.80--0.97) & 0.36(0.23--0.63) & 5.0(3.8--6.6)                       
   & 3.6(3.3--3.8)    & 0.022 (fixed) & 281/320\\
gv & 0.90(0.82--0.98) & 0.31(0.21--0.50) & 6.5(5.2--9.1)                       
   & 1.7(1.5--1.9)    & 0.44(0.17--0.86) & 279/319\\
\sidehead{\bf NGC 4649:}
gx & 0.68(0.56--0.83) & 0.80(0.35--1.2) & 4.2(3.9--4.6)                       
   & ...              & ...              & 271/247\\
gp & 0.81(0.67--0.93) & 0.27(0.16--0.49) & 15(8.6--26)                       
   & 3.9(3.4--4.4)    & 7.1(4.6--9.4)    & 255/246\\
gv & 0.75(0.65--0.91) & 0.27(0.18--0.58) & 9.6(6.7--19)                       
   & 3.4(3.0--3.8)    & 0($<0.037$)      & 264/246\\
\enddata

\tablecomments{All models include galactic (g) emission, absorbed by
the Galactic column density, consisting of interstellar thermal plasma
(temperature $kT_{\rm ISM}$ in keV, metallicity $Z_{\rm ISM}$ in solar
units) and X-ray binary bremsstrahlung (temperature $kT_{\rm bin}$)
components. In gx model fits, the fluxes of the two g-components are
independently varied for the two detectors; in gp and gv they are tied
together but an additional component is included for the PCA
(top-layer, 2.5-12.5 keV for NGC 4472 and 3-15 keV for NGC 4649)
spectrum -- an absorbed power-law (photon index $\Gamma$, column
$N_{\rm H,\Gamma}$) for model gp, a (Virgo) intracluster thermal
plasma (temperature $kT_{\rm ICM}$, metallicity $Z_{\rm ICM}$) for
model gv. Column densties are in units of $10^{22}$ cm$^{-2}$. Errors
correspond to $\Delta\chi^2=2.7$.}

\end{deluxetable}

\clearpage

\begin{deluxetable}{ccccccccc}
\tablecaption{PCA/GIS Joint Spectral Fits -- Fluxes}
\tablewidth{0pt}
\tablehead{
\colhead{Model} & \colhead{$f_{\rm ISM}$--GIS} & 
\colhead{$f_{\rm ISM}$--PCA} & \colhead{$f_{\rm bin}$--GIS} & 
\colhead{$f_{\rm bin}$--PCA} & \colhead{$f_{\rm ext}$--PCA} 
}
\startdata
\sidehead{\bf NGC 4472}
gx & 8.8(8.3--9.3) & 35(29--40) & 2.8(2.4--3.1) & 5.7(5.4--6.0)
   & ... \\
gp & 9.2(8.6--9.7) & 9.2(8.6--9.7) & 2.8(2.5--3.2) & 2.8(2.5--3.2)
   & 5.3(5.0--5.6) \\
gv & 9.4(8.9--9.9) & 9.4(8.9--9.9) & 2.8(2.6--3.2) & 2.8(2.6--3.2)
   & 5.0(4.7--5.3) \\
\sidehead{\bf NGC 4649}
gx & 3.6(3.3--3.9) & 4.5(0--19) & 1.4(1.2--1.5) & 6.2(6.1--6.3)
   & ... \\
gp & 3.3(3.1--3.6) & 3.3(3.1--3.6) & 1.8(1.6--2.0) & 1.8(1.6--2.0)
   & 4.1(4.0--4.3) \\
gv & 3.6(3.3--3.9) & 3.6(3.3--3.9) & 1.4(1.3--1.6) & 1.4(1.3--1.6) 
   & 4.8(4.6--5.0) \\
\enddata

\tablecomments{For detailed model explanations, see Table 1.  Fluxes
are in units of $10^{-12}$ erg cm$^{-2}$ s$^{-1}$ and are in the
0.5-2.0 keV band for the interstellar plasma component ($f_{\rm
ISM}$), and in the 2-10 keV band for the X-ray binary component
($f_{\rm bin}$) and $f_{\rm ext}$, where $f_{\rm ext}$ corresponds to
the extended power-law (in model gp) or thermal (in model gv)
component. For models gp or gv, the 2-10 keV interstellar component
flux is $\sim 1$ (0.2) $10^{-12}$ erg cm$^{-2}$ s$^{-1}$ for NGC 4472
(4649).}

\end{deluxetable}

\clearpage

\begin{deluxetable}{cccccccc}
\tablecaption{Diffuse 2-10 keV Surface Brightness}
\tablewidth{0pt}
\tablehead{
\colhead{Galaxy} & \colhead{PSPC} & \colhead{GIS} & \colhead{PCA--gv} & 
\colhead{PCA--gp} 
}
\startdata
NGC 4472 & 8.7(6.5--9.8) & 14.3(11.9--16.6) & 13.8(13.0--14.6) & 
14.7(13.8--15.5) \\
NGC 4649 & 7.3(5.5--8.2) & 15.3(11.5--18.5) & 13.3(12.8--13.8) & 
11.4(11.0--11.9) \\
\enddata

\tablecomments{Surface brightnesses are in units of $10^{-16}$ erg
s$^{-1}$ cm$^{-2}$ arcmin$^{-2}$ and are calculated for the GIS and
PSPC using the best-fit model for NGC 4472 from Table 1 that includes
a thermal extended third component (model gv).}

\end{deluxetable}

\clearpage


\begin{figure}
\figurenum{1a}
\centerline{\includegraphics[scale=0.70,angle=-90]{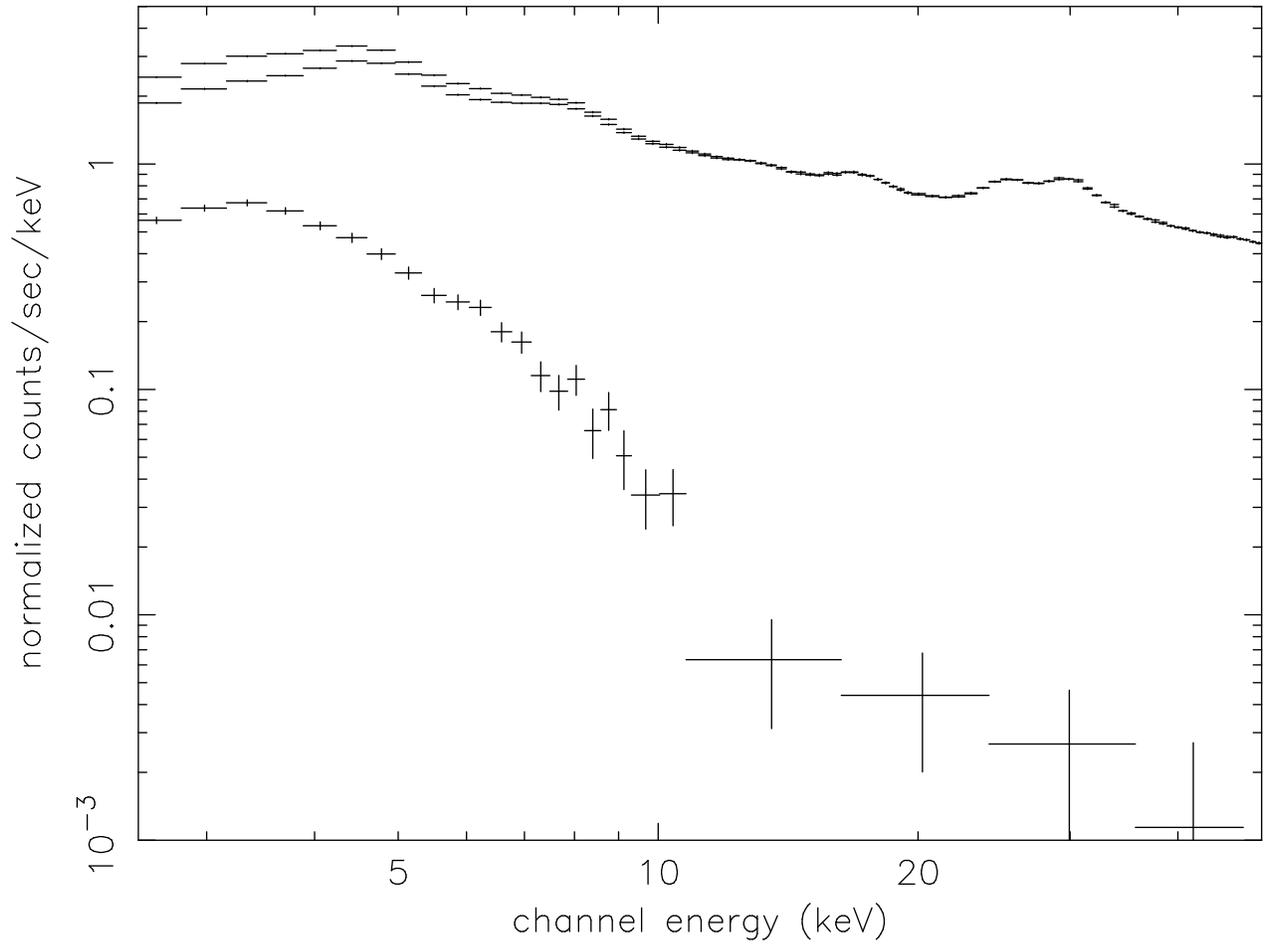}}
\caption{From top to bottom, total spectrum, model background
spectrum, and (background-subtracted) source spectrum for NGC 4472.}
\end{figure}
\clearpage

\begin{figure}
\figurenum{1b}
\centerline{\includegraphics[scale=0.70,angle=-90]{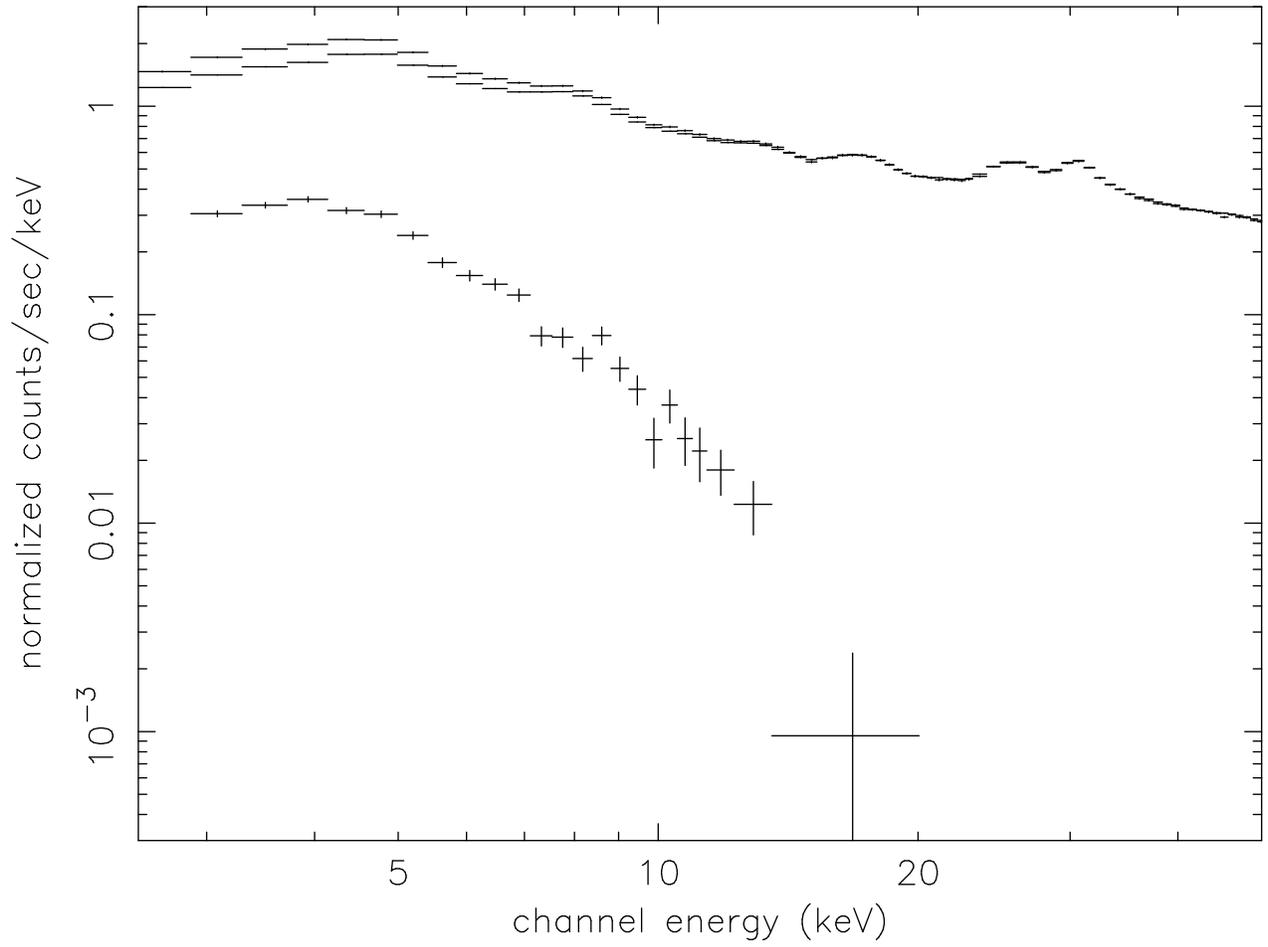}}
\caption{Same as Figure 1a for NGC 4649.}
\end{figure}
\clearpage

\begin{figure}
\figurenum{2}
\centerline{\includegraphics[scale=0.70,angle=-90]{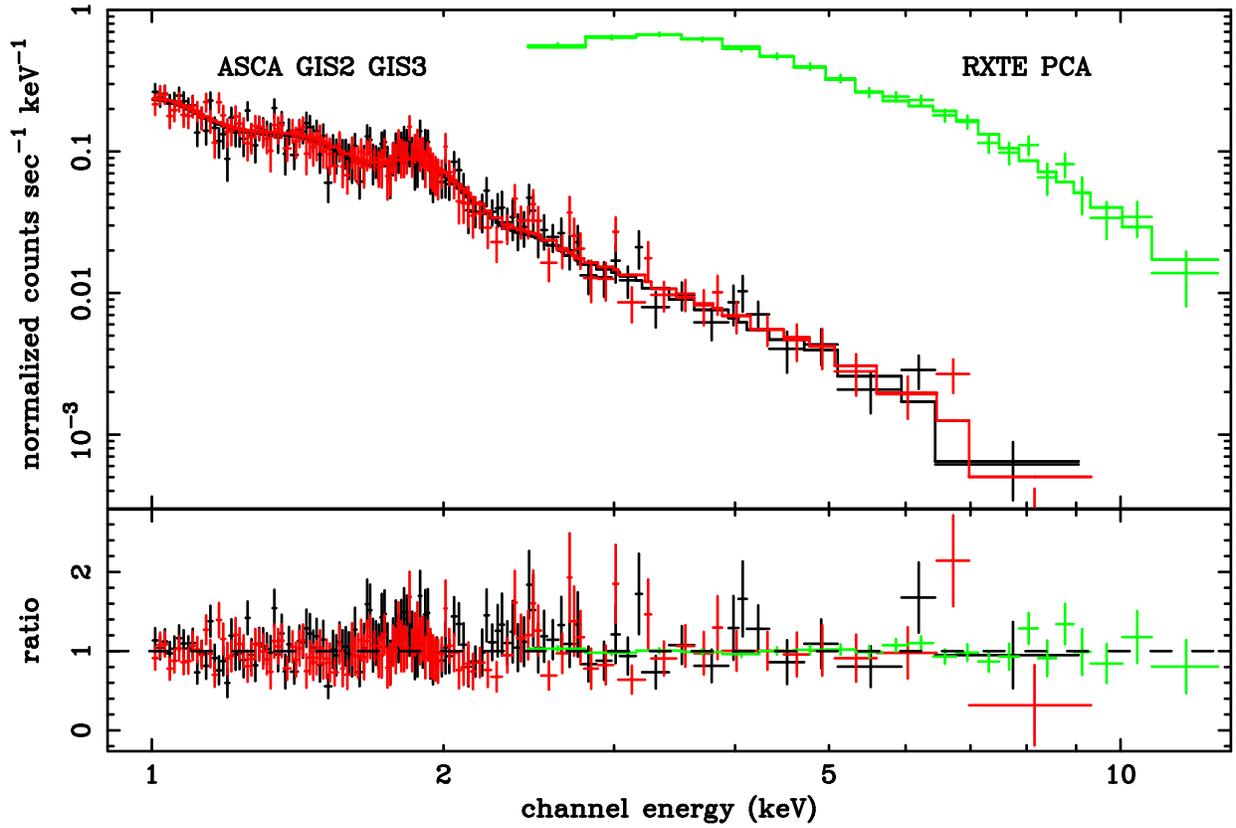}}
\caption{(top) {\it RXTE} PCA and {\it ASCA} GIS spectra $f(E)$
(errorbars) and best simultaneous fit model including a thermal
extended component (model gv in Tables 1 and 2) to joint dataset for
NGC 4472 (histogram). (bottom) Ratio of data to best-fit model.}
\end{figure}
\clearpage

\begin{figure}
\figurenum{3}
\centerline{\includegraphics[scale=0.70,angle=-90]{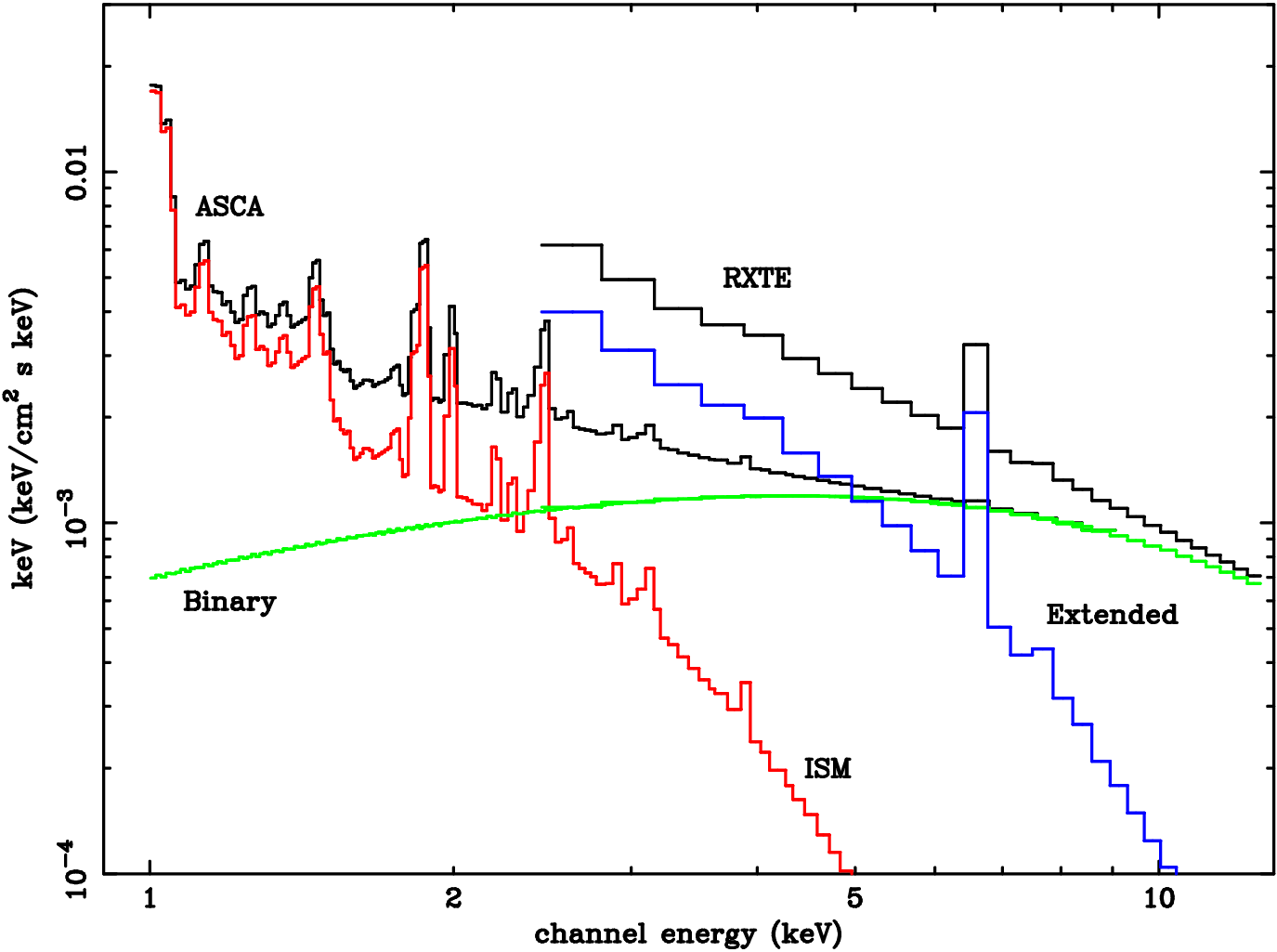}}
\caption{Best fit model including a thermal extended component
(histogram) for NGC 4472. The components are as follows: 0.9 keV
thermal plasma and 6.5 keV thermal bremsstrahlung models that
contribute to both spectra, and 1.7 keV thermal plasma that
contributes to the PCA spectrum only.}
\end{figure}
\clearpage

\begin{figure}
\figurenum{4}
\centerline{\includegraphics[scale=0.70,angle=-90]{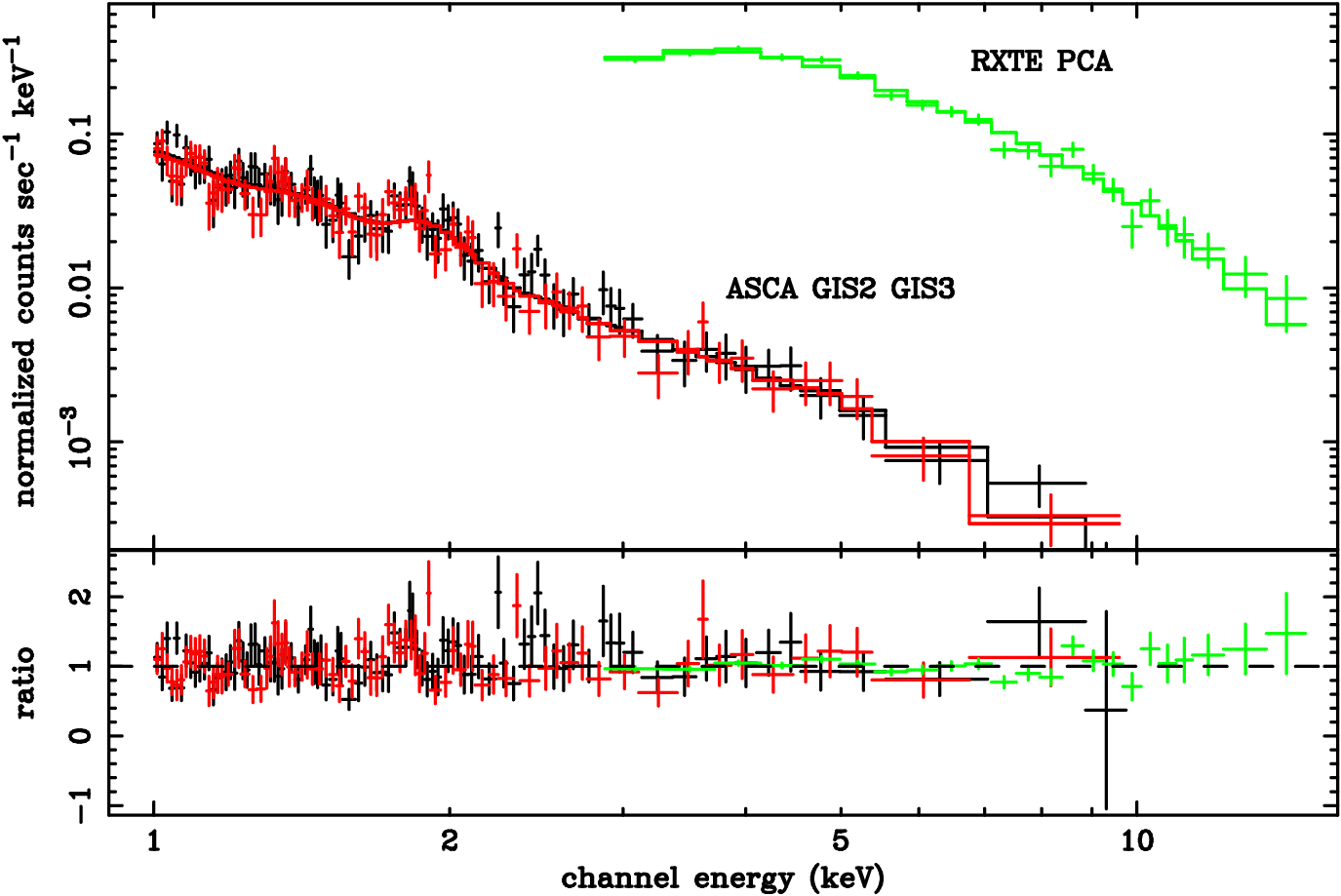}}
\caption{Same as Figure 2 for NGC 4649.}
\end{figure}
\clearpage

\begin{figure}
\figurenum{5}
\centerline{\includegraphics[scale=0.70,angle=-90]{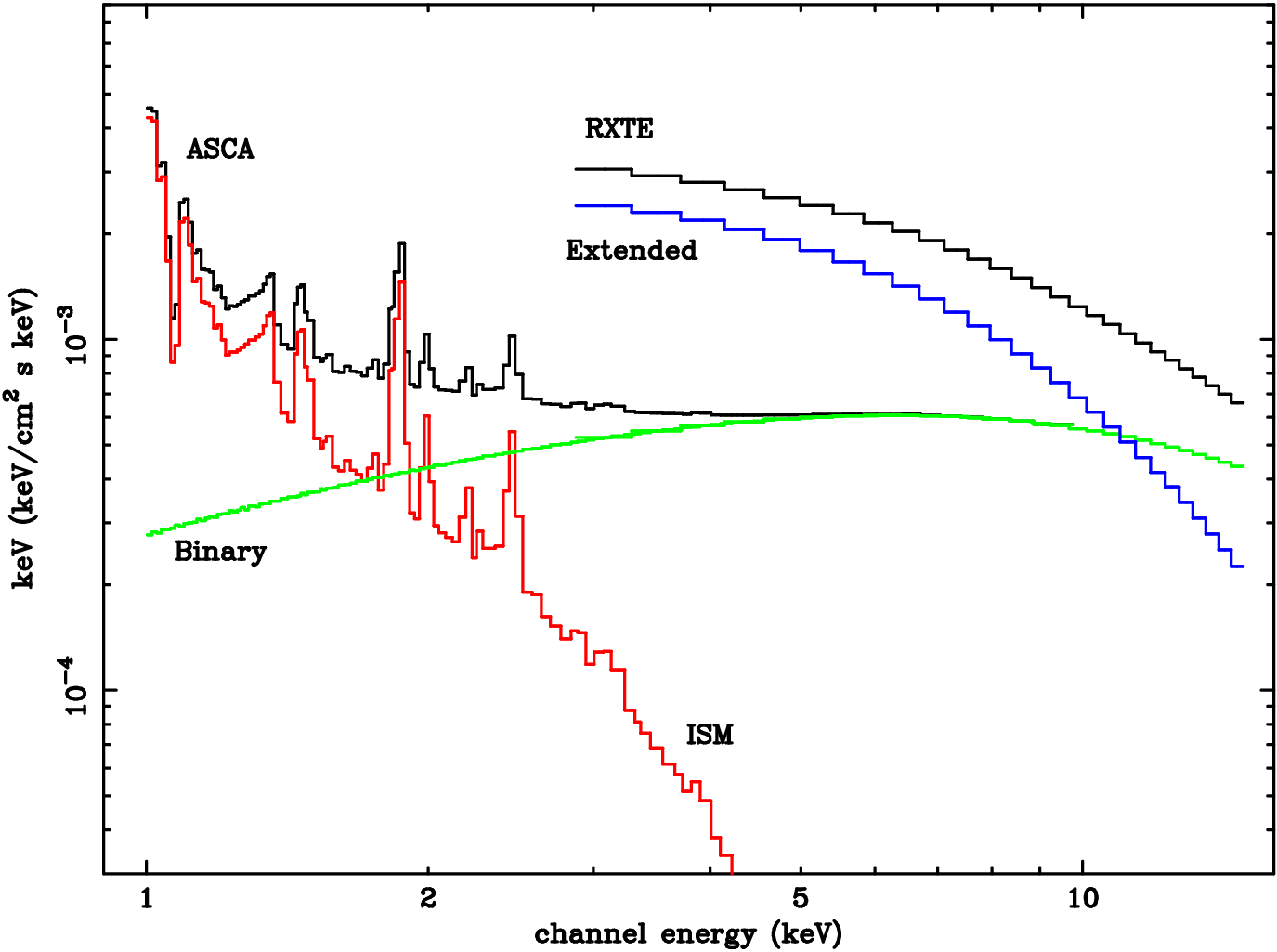}}
\caption{Best fit model including a thermal extended component
(histogram) for NGC 4649. The components are as follows: 0.75 keV
thermal plasma and 9.6 keV thermal bremsstrahlung models that
contribute to both spectra, and 3.4 keV thermal plasma that
contributes to the PCA spectrum only.}
\end{figure}
\clearpage

\begin{figure}
\figurenum{6}
\centerline{\includegraphics[scale=0.70,angle=-90]{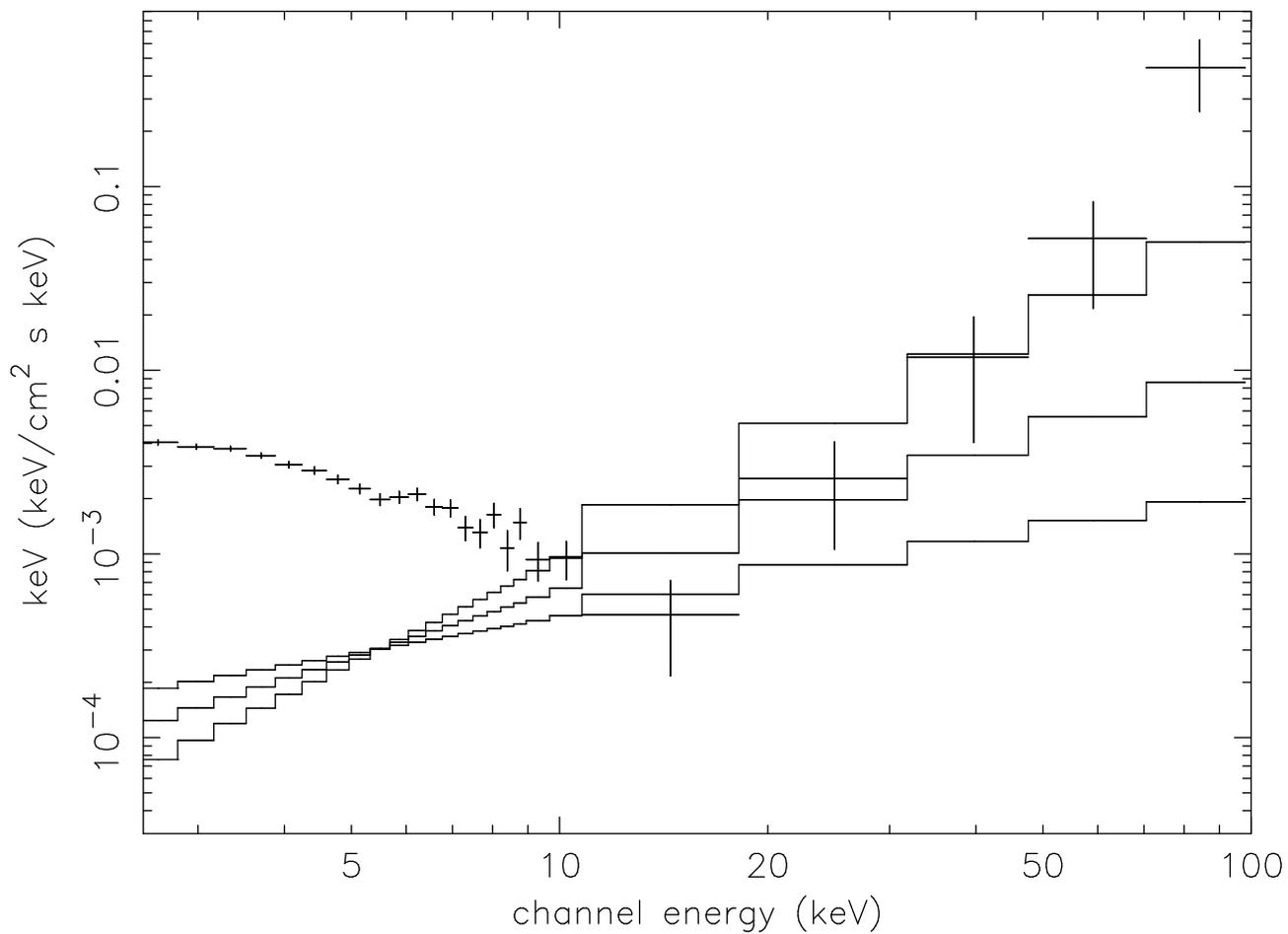}}
\caption{Unfolded NGC 4472 PCA spectrum $E^2f(E)$ (errorbars), and
power-law component (histograms: best-fit (middle) and 90\% confidence
upper and lower limits) from {\it ASCA} spectral decomposition of
Allen et al. (2000)}
\end{figure}
\clearpage

\begin{figure}
\figurenum{7}
\centerline{\includegraphics[scale=0.70,angle=-90]{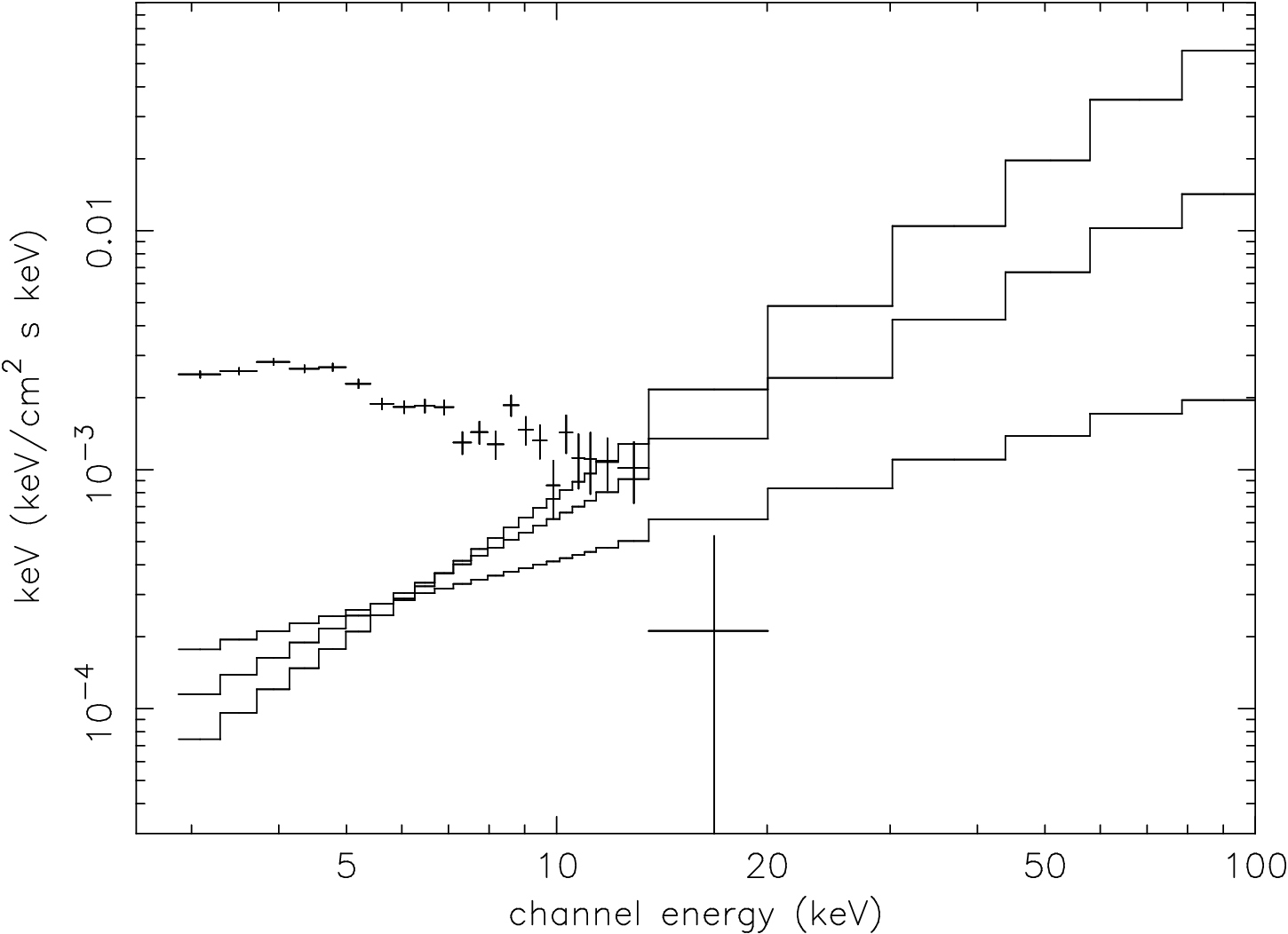}}
\caption{Same as Figure 6 for NGC 4649.}
\end{figure}
\clearpage

\end{document}